\documentclass[final]{elsart}

\usepackage{graphicx}

\usepackage{amssymb}
\usepackage{url}

\newcommand{\psp}{\emph{PulsarSpectrum}}
\newcommand{\til}{~}
\newcommand{\nat}{Nature}
\newcommand{\sci}{Science}
\newcommand{\apjl}{ApJL}
\newcommand{\apj}{ApJ}
\newcommand{\aj}{AJ}

\newcommand{\aap}{A\&A}
\newcommand{\aaps}{A\&AS}
\newcommand{\apss}{Ap\&SS}
\newcommand{\apjs}{ApJS}
\newcommand{\mnras}{MNRAS}

\newcommand{\phrl}{PhRvL}

\begin{document}
\maketitle

\begin{frontmatter}


\title{Pulsar Simulations for the \emph{Fermi} Large Area Telescope}

\author[infnpi]{M. Razzano\corauthref{cor}},
\ead{massimiliano.razzano@pi.infn.it}
\author[gsfc]{A.K. Harding\corauthref{cor}}
\corauth[cor]{Corresponding author.},
\ead{Alice.K.Harding@nasa.gov}
\author[infnpi]{L. Baldini},
\author[infnpi]{R. Bellazzini},
\author[infnpi]{J. Bregeon},
\author[uniwa]{T. Burnett}
\author[slac]{J. Chiang},
\author[kipac,slac]{S. W. Digel},
\author[slac]{R. Dubois},
\author[infnpi]{M. W. Kuss},
\author[infnpi]{L. Latronico},
\author[gsfc]{J. E. McEnery},
\author[infnpi]{N. Omodei},
\author[infnpi]{M. Pesce-Rollins},
\author[infnpi]{C. Sgr\`o},
\author[infnpi]{G. Spandre},
\author[gsfc]{D. J. Thompson}

\address[infnpi]{Istituto Nazionale di Fisica Nucleare sez. Pisa, Largo B. Pontecorvo 3, I-56127 Pisa,Italy}
\address[gsfc]{Astrophysics Science Division, Code 661, NASA Goddard Space Flight Center, Greenbelt, MD 20771, USA}
\address[gsfc1]{Astrophysics Science Division, Code 663, NASA Goddard Space Flight Center, Greenbelt, MD 20771, USA}
\address[kipac]{Kavli Institute for Particle Astrophysics and Cosmology, Stanford University, Stanford, CA 94309}
\address[slac]{Stanford Linear Accelerator Center, 2575 Sand Hill Road, Menlo Park, CA 94025, USA}
\address[uniwa]{Department of Physics, University of Washington, Seattle WA, 98195-1560 USA}

\vspace{2cm}
Contact authors:\\
\begin{itemize}
	\item Massimiliano Razzano. \emph{Address:} Istituto Nazionale di Fisica Nucleare sez. Pisa, Largo B. Pontecorvo 3, I-56127 Pisa, Italy. ( email address: massimiliano.razzano@pi.infn.it )
	\item Alice K. Harding. \emph{Address:} Astrophysics Science Division, Code 663, NASA Goddard Space Flight Center, Greenbelt, MD 20771, USA. ( email address: Alice.K.Harding@nasa.gov )
\end{itemize}


\newpage
\begin{abstract}
Pulsars are among the prime targets for the Large Area Telescope (LAT) aboard the recently launched $Fermi$ observatory. The LAT will study the gamma-ray Universe between 20 MeV and 300 GeV with
unprecedented detail. Increasing numbers of gamma-ray pulsars are being firmly identified, yet 
their emission mechanisms are far from being understood.
To better investigate and exploit the LAT capabilities for pulsar science, a set of new detailed pulsar simulation tools have
been developed within the LAT collaboration. The structure of the pulsar simulator package (\psp) is presented here.
Starting from photon distributions in energy and phase obtained from theoretical calculations or phenomenological  considerations,
gamma rays are generated and their arrival times at the spacecraft are determined  by taking into account effects such as barycentric effects and timing noise. Pulsars in binary systems also can be simulated given orbital parameters.
We present how simulations can be used for generating a realistic set of gamma rays as observed by the LAT, focusing on some case studies that show the performance of the LAT for pulsar observations.
\end{abstract}

\end{frontmatter}

\section{Introduction}\label{sec:intro}
Pulsars are among the most intriguing sources in the gamma-ray sky and represent unique laboratories for probing physical laws
in extreme environments. The EGRET experiment aboard the Compton Gamma Ray Observatory (CGRO), whose mission
ended in 2000, made the first all-sky survey above 50 MeV and made breakthrough observations of high-energy gamma-ray
pulsars \cite{thompson93}. CGRO observed the already-known Vela and Crab pulsars, discovered five more gamma-ray pulsars, including the pulsed detection of Geminga, which at the time had been known just as a point source. 
Yet, many sources in the 3$^{rd}$ EGRET catalog
 are unidentified \cite{hartman99}, and pulsars are believed to account for the majority of unidentified sources near the Galactic plane \cite{yadiga95,gonthier02}. The $Fermi$ Gamma-ray Space Telescope, a new international and multi-agency space observatory, will be able to identify many of them unambiguously by searching for periodicity with high accuracy in space and time.\\
In spite of the efforts to understand the emission mechanisms in pulsars many questions still remain unanswered. 
A more detailed knowledge of high-energy gamma-ray emission from pulsars will help
 to unveil their real natures and the $Fermi$ observatory, successfully launched aboard a Delta II rocket from Cape Canaveral Air Force
 Station on 11 June 2008, will provide deep insights into these fascinating objects.\\
$Fermi$, formerly known as GLAST, carries two main instruments: the Large Area Telescope (LAT) which is a pair conversion telescope
for gamma rays with energies between 20 MeV and 300 GeV, and the Gamma-ray Burst Monitor (GBM), devoted to the observation of Gamma Ray
Bursts.\\
The LAT consists of a high-precision silicon strip Tracker, a hodoscopic CsI Calorimeter, and a
segmented Anticoicidence Detector which covers the full Tracker.  A comprehensive description of the LAT can
be found in \cite{atwood09}.\\  $Fermi$ will operate primarily in scanning mode, pointing away from the Earth and rocking about
the zenith direction. The entire sky will be viewed every two orbits ($\sim$ 3 hours), providing long and uniform coverage.
The high sensitivity due to the large effective area ($\sim$ 3800 cm$^{2}$ at E $\sim$ 100 MeV and $\sim$ 8000 cm$^{2}$ for E$>$1GeV on-axis), the sharp Point Spread Function ($\theta_{68}\sim$0.1$^{\circ}$ for E$>$1 GeV), the large field of view (2.4 sr) will allow the LAT to detect source fluxes down to $\sim 3\times$10$^{-9}$ ph cm$^{-2}$s$^{-1}$ \footnote{for a steady source at E $>$100 MeV after 1 year sky survey and assuming an E$^{-2.1}$ spectrum with no spectral cutoff and a high latitude diffuse flux of 1.5 $\times$ 10$^{-5}$ ph cm$^{-2}$s$^{-1}$sr$^{-1}$\cite{atwood09}}. This corresponds to a sensitivity $\sim$ 30 greater than
EGRET's.
An important factor for pulsar studies is the absolute timing accuracy of the LAT, which will be $<$10 $\mu$s for each detected
gamma ray. $Fermi$ has detected the EGRET pulsars and and so far detections of two new ones, the radio quiet pulsar in the CTA 1 supernova remnant \cite{abdo08a} and PSR J2021+3651 \cite{abdo08b}, which has been also independently discovered by AGILE mission \cite{halpern08}.\\
To better understand the high-level performance of the LAT with regard to pulsar astronomy, a complete, new powerful simulation
package (\psp) has been developed.\\
With respect to the simulation capabilities of existing packages, such as TEMPO2 \cite{hobbs06} which have been used
successfully for pulsar analysis by radio astronomers, \psp\til is more focused on specific issues of interest in gamma-ray astronomy, as for example the possibility to include more complex spectra.
Moreover, this simulator has the advantage of being specific for gamma-ray observations and fully compatible with the software that simulates the response of the LAT.\\
The pulsar simulation package was a very useful tool before the launch of $Fermi$ and it will continue to be so, for
comparing Monte Carlo results with real data, making it easier to constrain the statistical significance of the results.\\
In pulsar simulations the most complex feature is to take into account any effects which can influence the photon arrival time
at the telescope. They depend mainly on the spacecraft motion, and on the change of period due to rotational energy losses. Moreover, pulsars do not show a perfectly-steady increase of the period but exhibit timing noise, which becomes particularly important for young gamma-ray pulsars. This noise significantly reduces the
phase coherence with time and represents an important limitation when searching for pulsations over long time periods.
\psp\til takes into account all these effects when computing the arrival times of the simulated gamma rays. It has been widely used by the LAT collaboration for full simulation of the expected pulsar population in the sky together with other sources.
It has also been included in the LAT \emph{Standard Analysis Environment} (SAE), that has been developed by the $Fermi$ Science Support Center\footnote{$http://fermi.gsfc.nasa.gov/ssc/$}.
\section{Gamma-ray pulsar simulation: motivations and basics}\label{sec:gpulsars}
In designing \psp\til all the observational information gathered so far about the known gamma-ray pulsars were used. EGRET discovered the gamma-ray pulsars PSR B1706-44 \cite{thompson92}, PSR B1055-52 \cite{fierro93} and PSR B1951+32
\cite{ramanamurthy95} by searching for gamma-ray pulsations at the same periods as their radio counterparts. Even though it was detected as a point source by SAS-2 and COS B, Geminga was not yet identified. The nature of Geminga as a gamma-ray
pulsar was finally revealed when EGRET found a gamma-ray periodicity of about 237 ms starting from the X-ray pulsations detected by
ROSAT \cite{halpern92} and identified Geminga as a radio-quiet pulsar\footnote{The Geminga pulsations were later found in COS B \cite{bignami92} and SAS-2 data \cite{mattox92} and the light curves were
compared with the EGRET one\cite{mattox98}} \cite{bertsch92}. In addition, the BATSE and COMPTEL instruments aboard CGRO also detected PSR B1509-58
\cite{ulmer93}, even though this pulsar was never detected at EGRET energies. More recently, $Fermi$ has discovered a new radio-quiet pulsar in the CTA 1 supernova remnant using blind search of periodicity in gamma rays \cite{abdo08a}.\\
The light curves of the EGRET detected pulsars are energy-dependent and above 100 MeV they show a double-peaked profile, while at  $>$~GeV
energies one peak is reduced, usually the first peak coming after the radio pulse\footnote{For all the simulations we adopt the convention of assigning phase 0 to the radio pulse.}.
The peaks of the observed gamma-ray pulsars can be modeled as Lorentz profiles (e.g. \cite{kanbach94}). This profile has been chosen
as a template when simulating random-generated light curves.\\
Spectral analysis shows that most of the power output comes from the hard X-ray and gamma-ray~ emission, the latter being modeled by a non-thermal spectrum \cite{thompson99}. EGRET spectral analysis of the brightest pulsars shows spectral breaks at GeV energies, and observations by ACTs of the brightest pulsars were unable to see pulsed emission \cite{aharonian07}, although recently MAGIC has observed pulsed gamma rays above 25 GeV from the Crab pulsar \cite{aliu08}. The EGRET spectral observations and the absence of pulsed emission by ACTs (with the exception of the Crab) suggests a spectral cutoff that provides an important
constraint on the main gamma-ray emission models proposed for pulsars \cite{harding01}. In Polar Cap models, emission takes place at low altitude ($<$ neutron star radius) above magnetic poles, and it is composed of emission from synchrotron-pair cascades that are initiated by curvature radiation converting via $\gamma-B \rightarrow e^{+}e^{-}$ absorption \cite{daugherty96}. Owing to the absorption in high B-fields, this model predicts sharp, super-exponential cutoffs in the observed spectra at $\sim$ few GeV energies (See Eq. \ref{eq:specndj}). 
An evolution of the Polar Cap models are the Slot Gap models \cite{muslimov04}, where particle acceleration takes place in thin slot gaps along the last open field line from the neutron star surface to the light cylinder. In the Outer Gap model \cite{romani96}, instead, particles are accelerated within the vacuum outer gaps and emit gamma rays far from the neutron star surface mainly by curvature and synchrotron radiation. In high-altitude Slot Gap and Outer Gap models a simple exponential cutoff is predicted, due to the radiation reaction limit of the accelerated particles.
\psp\til has great flexibility for simulating different spectral shapes according to these theoretical scenarios, since a power law spectrum with cutoff can be simulated in \psp\til and its energy and shape can be adjusted according to the model. Thanks to this feature it will be possible to compare the distribution of expected photons with real data from the LAT, in order to constrain the theoretical emission scenarios.\\
Phase-resolved spectral analysis of EGRET observations of Vela, Geminga and Crab pulsars show that no simple pattern can be easily
recognized. However a trend in phase-dependent spectra has been found: the peaks appear softer than inter-peak emission and harder
than the outer wings \cite{nolan96}. The dependence of the spectrum on phase is thus an important feature which needs to be simulated
using a specific model in \psp.\\
In order to give some numerical estimates on the relative count rates obtained from simulations at different energy ranges, we used the Vela pulsar as an example, obtaining for E $>$ 100 MeV a daily rate of $\sim$ 440 counts from the pulsar, $\sim$ 60 counts from the Galactic and extragalactic gamma-ray diffuse emission, $\sim$ 5 counts from residual charged background and an upper limit of 10 counts from the PWN \cite{abdo08c}. At highest energies (E $>$ 1 GeV), we obtained a daily rate of $\sim$ 140 counts from the pulsar and $\sim$ 16 counts from the Galactic gamma-ray diffuse, while the extragalactic diffuse and residual background give a negligible contribution.
\section{\emph{PulsarSpectrum} overview}\label{sec:pspover}
\psp\til generates a bi-dimensional gamma-ray photon histogram $N_{\nu}$(E$_{i}$,t$_{i}$)\footnote{In the following $N_{\nu}$(E$_{i}$,t$_{i}$)
will be abbreviated with $N_{\nu}$.}, which represents the differential photon flux as a function of 
energy bins $E_{i}$ and time bins $t_{i}$ expressed in ph m$^{-2}$ s$^{-1}$. $N_{\nu}$ is used to extract energies and time intervals between the simulated photons,
subsequently corrected for timing effects (Sec. \ref{sec:teffects}).\\
\psp\til has two simulation strategies for creating $N_{\nu}$:
one is based on phenomenological considerations and uses analytical modeling of the pulsar spectrum;
the other one is suited for simulations based on more
complex pulsar spectra, e.g. phase-dependent spectra that cannot be modeled analytically.\\
After $N_{\nu}$ is created, \psp\til calculates the time interval between subsequent photons according to the source
flux. Each generated photon is sent to the Monte Carlo simulation of the LAT
for the reconstruction.
\subsection{The phenomenological model}
\begin{figure}[ht]
\begin{center}
\includegraphics[width=0.6\columnwidth]{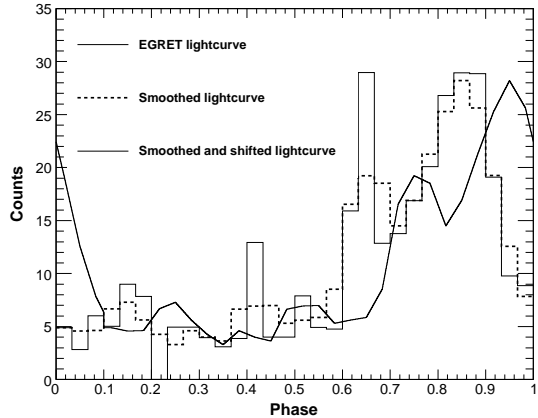}
\caption{Example of creation of a simulated light curve based on EGRET observation of PSR B1055-52 using the phenomenological model in \psp.
The solid histogram reproduces the EGRET light curve from \cite{thompson99}. The dotted histogram is the result of a boxcar
smoothing with 3-bins width. The solid curve is the smoothed histogram with increased number of bins and a phase shift of $\Delta \phi$=0.1 to assign the phase 0 to the first bin.}\label{fig:fig1_1055m52}
\end{center}
\end{figure}

This model creates an $N_{\nu}$
flux histogram by multiplying a pulsar light curve and spectrum derived from phenomenology.
The light curve can be randomly generated with one or two Lorentzian peaks or loaded from a file that
lists the EGRET counts for each phase bin. Although a more detailed model should use the flux instead of counts, 
these are available only for the brightest EGRET pulsars \cite{fierro98}; the light curves obtained are realistic enough for simulations.\\
LAT light curves of faint EGRET pulsars are produced first by applying a smoothing boxcar filter to the EGRET
light curves in order to reduce the statistical fluctuations, then by increasing the number of bins through linear interpolation
to get time bin width of $\sim$10 $\mu$s, in accordance with the LAT requirements for absolute
time resolution \cite{smith08}. An example of this algorithm applied to the simulation of PSR B1055-52 is shown in
Fig. \ref{fig:fig1_1055m52}.
Pulsar spectra are built by using the analytical formula of a power law with a super
exponential cutoff \cite{nel95}. This is a good model for the phase-averaged gamma-ray pulsar spectrum and allows for an adjustable shape of the energy cutoff.\\
The parametric expression of the differential spectrum is:
\begin{equation}\label{eq:specndj}
\frac{dN}{dEdAdt} \propto \left(\frac{E}{E_{n}}\right)^{-\alpha}
\exp\left[-\left(\frac{E}{E_0}\right)^{b}\right]
\end{equation}
where $\alpha$ is the power law spectral index, $E_{0}$ the cutoff energy, $b$ the index which models
the shape of the cutoff, and $E_{n}$ a scale factor set to 1 GeV for the standard simulations. These parameters are user selectable.\\
Eq.\ref{eq:specndj} is useful for modeling the exponential cutoff ($b$=1) predicted by outer magnetosphere emission models or the
super-exponential cutoff ($b>$1) predicted by the Polar Cap emission model.
The spectrum of Eq.\ref{eq:specndj} is multiplied by the light curve $T(\phi)$ in order to obtain $N_{\nu}$:
\begin{equation}\label{eq:dflux}
N_{\nu}(E,t) = C\frac{dN}{dEdSdt}(E)T(\phi)= C\frac{dN}{dEdSdt}(E)T(\frac{t}{P_{0}}) \til \til \mbox{(ph m$^{-2}$ s$^{-1}$)}
\end{equation}
where the rotational phases $\phi$ have been converted to times $t$ using the pulsar period $P_{0}$\footnote{For compatibility with the other source simulators developed by the LAT the time is used instead of the rotational phase.}.
The normalization factor $C$ is computed in order to obtain an integrated flux above 100 MeV equal to the total flux of the pulsar
in the EGRET energy band (100 MeV - 30 GeV) given in input. This model has been extensively used to check pulsar analysis tools and perform basic sensitivity studies \cite{razzano07}.
\subsection{The case of complex emission patterns}
\begin{figure}[ht]
\begin{center}
\includegraphics[width=0.6\columnwidth]{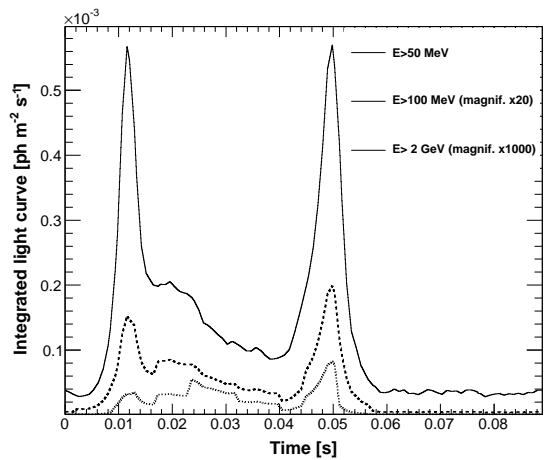}
\caption{Light curves for the Vela pulsar at 3 energy ranges obtained from the phase-dependent spectral model based on
EGRET observations. The spectral cutoff has been set to 8 GeV for the whole phase interval. The light curves for E$>$100 MeV and E$>$2 GeV have been magnified in order to compare the profiles in
the same plot.}\label{fig:fig2_velaelc}
\end{center}
\end{figure}

For simulation of more complicated phase-energy distributions, another approach must be used. The basic idea is simple: \psp\til loads an external $N_{\nu}$ bi-dimensional histogram and normalizes it according to the flux.
In this way the computation of $N_{\nu}$ is delegated to external packages so many more scenarios
can be used to produce a customized pulsar simulation.
This kind of simulation is very useful to understand the utility of the LAT data for phase-resolved spectroscopy.
As an example, we prepared a phase-dependent spectrum of the Vela pulsar based on EGRET observations \cite{fierro98}. A super-exponential cutoff has been applied for the peak phase intervals and an exponential cutoff for the remaining phase intervals. From this model spectrum it is possible to estimate how the modeled light curve varies with energy (Fig. \ref{fig:fig2_velaelc}). This model reproduces the behavior observed by EGRET, since the first peak is strongly reduced at high energies, leaving only the second one \cite{thompson05}.
\subsection{Timing corrections}\label{sec:teffects}
The $N_{\nu}$ histogram is used to produce a list of photons, each
one with assigned energy, arrival direction (from the pulsar position) and rotational phase. The simulator then calculates the arrival times as the LAT would measure them. In order to compute these arrival times, timing corrections must be applied so that after processing the simulated photons with the LAT pulsar tools the original light curve can be recovered. In the first stage the time intervals between photons are computed by assuming that:
    \begin{enumerate}
     \item the pulsar period is constant in time,
     \item the LAT is fixed with respect to the pulsar,
     \item there is no timing noise and
     \item the pulsar is not in a binary system.
   \end{enumerate}
 In this approximation the time intervals can be calculated considering the flux of the source and Poisson statistics. For fluxes comparable to those the gamma-ray pulsars, the time intervals computed by \psp\til do not
significantly differ from those computed for a pure Poisson process. The simulator assume that the flux is constant with time, as observed from EGRET observations \cite{kanbach94}.
The assumptions listed above lead to simplified calculations but do not provide realistic time intervals $\delta t$ and so
corrections must be applied. In order to provide a realistic list of pulsar photons several effects must be incorporated as corrections to $\delta t$, that are described below.
\subsubsection{Barycentric effects}\label{sec:barydecorr}
Since the Earth, and the $Fermi$ observatory orbiting around it, undergo significant acceleration with respect to the pulsar, the first step in analyzing pulsar data is to convert the arrival times recorded at the spacecraft to the Solar System Barycenter (SSB) reference frame, which is nearly inertial with respect to the pulsar.\\
Photon arrival times at the spacecraft are expressed in MET\footnote{Mission Elapsed Time. For $Fermi$ is defined as the number of seconds since 2001 January 01 at 00:00:00 (UTC).}Terrestrial Dynamical Time (TT or TDT).
  The relation between the arrival time t$_{B}$, expressed in Barycentric Dynamical Time TDB\footnote{According to IAU
 Standards the Barycentric Dynamical Time is the independent variable of the equations of motion with respect to the barycenter
  of the Solar System. It is related to TT by a mathematical expression that includes positions of Solar System bodies
  \cite{hobbs06}.}, and TT arrival time t$_{F}$ at the spacecraft is:
\begin{equation}\label{eq:bcorr}
t_{B} = t_{F} + \Delta t_{C} (t_{F}) + \Delta t_{E} (t_{F}) + \Delta t_{R} (t_{F}) + \Delta t_{S}(t_{F})
\end{equation}
where $\Delta t_{C} (t_{F})$ is the \emph{clock correction}, $\Delta t_{E} (t_{F})$ is the \emph{Einstein delay}, $\Delta t_{R} (t_{F})$ is the \emph{Roemer delay} and $\Delta t_{S} (t_{F})$ is the \emph{Shapiro delay}. Starting from $t_{B}$, the inverse of Eq. \ref{eq:bcorr}
is computed in order to give the arrival times $t_{F}$ at the spacecraft.\\
$\Delta t_{C} (t_{F})$ is the correction from recorded time at the spacecraft to International Atomic Time and then to TT\footnote{According to IAU Standards Terrestrial Time TT is the time reference for apparent geocentric ephemerides and
 it is related to the \emph{International Atomic Time} (TAI) as as TT = TAI + 32.184 s \cite{hobbs06}.}. The \emph{Einstein delay} $\Delta t_{E} (t_{F})$ is a relativistic correction based on the variation in clock rate at the spacecraft and in the SSB because of the motion of the Earth and its gravitational potential. The values of $\Delta t_{C} (t_{F})$ and 
$\Delta t_{E} (t_{F})$ are of $\sim$1.5 $\mu$s and $\sim$1.6 ms respectively \cite{hobbs06} and are computed from the JPL ephemerides\footnote{It is possible in \psp\til to use the Jet Propulsion Laboratory DE200 or DE405 ephemerides \cite{standish98}.} using the standard routine AXBARY.C\footnote{See $http://heasarc.gsfc.nasa.gov/docs/software/lheasoft/RelNotes04.html$.}. The \emph{Roemer delay} $\Delta$t$_{R}$(t$_{F}$) is due to the light propagation time from the position of $Fermi$
to the SSB and it is based on the location of the Earth, the Sun and the $Fermi$ spacecraft, and can reach $\sim$500 s of amplitude. It can be written as:
\begin{equation}
\Delta t_{R}(t_{F}) = -\frac{1}{c}\left[ \mathbf{r}_{FE}(t_{F}) + \mathbf{r}_{ES}(t_{F}) + \mathbf{r}_{SB}(t_{F})\right]\cdot \mathbf{\hat{s}}
\end{equation}
where \textbf{r$_{FE}$} is the vector from $Fermi$ spacecraft to the center of the Earth, \textbf{r$_{ES}$} from the center of
Earth to the center of the Sun and \textbf{r$_{SB}$} from the center of the Sun to the SSB.
$\textbf{$\hat{s}$}$ is the unit vector identifying the sky position of the pulsar, and can be considered constant, since proper motion and
parallax effects are neglected by \psp.The \emph{Shapiro delay} $\Delta t_{S}$(t$_{F}$) is a relativistic correction due to the gravitational
field of the Sun \cite{shapiro64}.  At the first order $\Delta t_{sh}(t_{F})$ can be calculated as:
\begin{equation}
\Delta t_{sh}(t_{F}) = \frac{2G}{c^{2}}\log(1-\mathbf{\hat{s}}\cdot \mathbf{\hat{r}}_{FS})
\end{equation}
where $\mathbf{r}_{FS}$ is the vector from $Fermi$ to the Sun and $G$ the gravitational constant. The Shapiro delay becomes significant only for sources that are located at small elongation from the Sun. Higher-order corrections are due to the Sun ($\sim$100$\mu$s), while corrections due to other planets ($<$180 ns) are neglected here \cite{hobbs06}. 
\subsubsection{Period changes}\label{sec:pchange}
Since the rotational energies of pulsars decrease with time, the periods increase. \psp\til accounts for this effect by correcting the arrival times. When building a gamma-ray pulsar
light curve the second step after barycentering is to assign  a rotational phase to each photon.
To do this, it is necessary to know the {\em spin parameters}, i.e. the rotational
frequency {\em f$(t_{0}$)} and its derivatives {\em $\dot{f} (t_{0})$} and {\em $\ddot{f} (t_{0})$}
at a particular epoch t$_{0}$.
In the inertial frame of the pulsar the rotational phase $\phi(t)$ can be written:
\begin{equation}\label{eq:phit}
\phi(t) = frac[\phi(t_{0})+f(t_{0})(t-t_{0}) + \frac{1}{2}\dot{f}
(t_{0})(t-t_{0})^{2} + \frac{1}{6}\ddot{f} (t_{0})(t-t_{0})^{3}].
\end{equation}
where \emph{frac} indicates the fractional part of the right hand member of the equation. It is also possible to express the spin parameters as the period $P_{0}$ and its derivatives $P_{1}$ and $P_{2}$.\\
Higher-order terms in Eq.\ref{eq:phit}, which better describe period evolution, are not believed to represent the secular spin down of the pulsar, but rather timing irregularities. For this reason they are included in the simulation of the timing noise.
\subsubsection{Timing Noise}\label{sec:tnoise}
\begin{figure}[ht]
\begin{center}
\includegraphics[width=1.0\columnwidth]{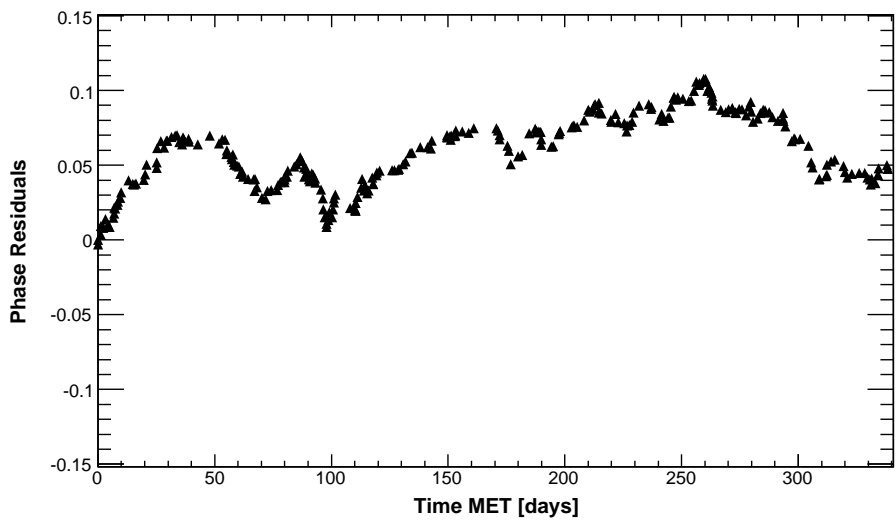}
\caption{Example of the Random Walk timing noise implemented in \psp. In this case a pure phase timing noise has been
applied to a simulated pulsar with period $P$ = 0.28 s and a $P_{1}$ = 1.1$\times10^{-13}$ s/s, with a corresponding
activity A$\approx$1.96.}\label{fig:fig3_pntn}
\end{center}
\end{figure}

For most of the pulsars the evolution of the period does not follow a steady linear increase and fluctuations in phase can occur, because of the so-called \emph{timing noise}. Since timing noise affects mainly young pulsars, it has a large impact on gamma-ray pulsars, which are mainly young objects. Thus a continuous monitoring of radio and X-ray counterparts of gamma-ray pulsar candidates is required \cite{smith08}. Moreover, 
it reduces the phase coherence over time, and limits the search of periodicity using only gamma rays in Radio Quiet pulsars (\emph{blind search})\cite{abdo08a}.\\
The rotational phase $\phi(t)$ at an arrival time $t$ can be modeled as a sum of a term $\phi_{S}(t)$, based on a steady increasing period (see Eq. \ref{eq:phit}), and a timing noise contribution $\phi_{N}(t)$. \psp\til implements a routine for computing $\phi_{N}(t)$ based on a Random Walk approach to timing noise \cite{cordes85}.\\
Within this approach we model a Random Walk in the $k$th derivative of the phase:
\begin{equation}\label{eq:rw}
\frac{d^{k}\phi(t)}{dt^{k}}=\sum_{i}a_{i}u(t-t_{i})
\end{equation}
where $u$ is the unit step function, $t_{i}$ are the times where discontinuous jumps in the $k$th derivative occur, and $a_{i}$ are the amplitudes of the jumps. The interval between the times $t_{i}$ of two subsequent jumps is governed by a Poisson distribution with mean rate $R$. Within this Random Walk model, $a_{i}$ is a random variable with zero mean and Root Mean Square (RMS) that can be estimated from the pulsar spin parameters. According to the model developed in \cite{cordes85}, \psp\til is able to calculate the RMS in case of pure \emph{Phase Noise} ($k$=0), pure \emph{Frequency Noise} ($k$=1), pure \emph{Slowing-down Noise} ($k$=2) or a linear superposition of them.\\
Eventually, the RMS can be related to $R$, to the time $T$ spanned by the data and to a term $\sigma _{TN}(2,T)$, which represents the RMS residual phase from a least-squares 2$^{nd}$ order polynomial fit over the interval $T$ (See Eq. 21 in \cite{cordes85}). Summarizing, the relations used by \psp\til to calculate the timing noise RMS are:
\begin{equation}\label{eq:s0}
\delta\phi^{2} \propto \frac{\sigma^{2}_{TN}(2,T)}{RT} \:\:\: ; \:\:\: \delta\nu^{2} \propto \frac{\sigma^{2} _{TN}(2,T)}{RT^{3}} \:\:\:;\:\:\: \delta\dot{\nu}^{2} \propto \frac{\sigma^{2}_{TN}(2,T)}{RT^{5}}
\end{equation}
In order to calculate the RMS for the different types of timing noise in Eq. \ref{eq:s0}, \psp\til uses $T$ as the arrival time of the photon, while the mean rate $R$=1 day$^{-1}$, in order to satisfy the validity condition $RT>$1 imposed by the model \cite{cordes85}. The remaining term $\sigma _{TN}(2,T)$ is computed using an
\emph{activity parameter} $A$, is defined as follows \cite{cordes85}:
\begin{equation}\label{eq:tna}
A = \log _{10}[ \sigma _{TN}(m,T)/\sigma _{TN}(m,T)_{Crab}]_{m=2}
\end{equation}
where proportionality coefficients are taken from \cite{cordes85}. $\sigma _{TN}(m,T)_{Crab}$ represents the RMS of the phase residuals for the Crab pulsar\footnote{The RMS of the residuals for the Crab pulsar in Eq. \ref{eq:tna} was chosen as the reference value because it was well-studied\cite{cordes85}.} and it is calculated as in Eq. 16 of \cite{cordes85}. The first step is therefore to calculate the activity parameter, which is done by \psp~ using the following relation \cite{cordes85}
\begin{equation}\label{eq:tna_pdot}
A = -1.37 + 0.71 \log P_{1}
\end{equation}
where the period derivative $P_{1}$ is expressed in units of 10$^{-15}$ s/s. Starting from the $\dot{P}$ of the simulated pulsar, \psp~ calculates $A$ (Eq. \ref{eq:tna_pdot}). This value is used to evaluate $\sigma_{TN}(m,T)$ (Eq. \ref{eq:tna}), and hence obtain $\delta\phi^{2}$, which represents the variance of the timing noise jumps (Eq. \ref{eq:s0}). Fig. \ref{fig:fig3_pntn} represents an example of simulated timing noise, which resembles the timing noise behavior shown in \cite{cordes85}. 
This timing noise model is only one of the possible ones. An interesting possibility for a future implementation of PulsarSpectrum is to introduce the phenomenological timing noise model obtained from TEMPO2 timing solutions on specific pulsar observations.
\subsubsection{Orbital modulation and binary corrections}\label{sec:binary}
Binary pulsars are an interesting target for the LAT.
About 7$\%$ of the known pulsars\footnote{Source: ATNF Pulsar Catalog($http://www.atnf.csiro.au/research/pulsar/psrcat$). Many simulations with \psp\til made use of the ATNF Pulsar Catalog\cite{manch05}. We would like to thank the ATNF team for maintaining this valuable resource.} and almost 80$\%$
of millisecond pulsars (MSPs) are observed in binary systems. Accretion
from a companion star in fact is believed to be responsible for the rejuvenation process that produces
millisecond pulsars.\\ 
In a binary system, the orbital motion and the gravitational field of the companion affect the timing of the pulsar and \psp\til can simulate these effects starting from the orbital parameters.
Dedicated routines for the correction of these effects are included in LAT SAE.
Orbits for non-relativistic binary systems can be described by Kepler's laws and calculations can be performed starting
from 6 \emph{Keplerian parameters} usually determined from observations at radio or other wavelengths\cite{kramer04}.\\
The parameters required by \psp\til to reference the photon arrival time to the barycenter of the binary system are the following:
\begin{itemize}
 \item orbital period $P_{b}$;
 \item projected semi-major axis $a_{p}\sin i$, where $i$ is the \emph{orbital inclination}, defined as the angle between
 the orbital plane and the plane of the sky;
 \item orbital eccentricity $e$;
 \item longitude of periastron $\omega$;
 \item epoch of periastron passage $T_{0}$;
 \item position angle of the ascending node $\Omega_{asc}$
\end{itemize}
To compute the binary correction it is necessary to solve the Kepler equation, which can be written using the
\emph{Eccentric Anomaly E(t)} in the following way:
\begin{equation}\label{eq:kepler}
E(t)-e\sin E(t) = \Omega_{b} \left[(t-T_{0})\right]	
\end{equation}
where $\Omega_{b}=2\pi/P_{b}$ is the mean angular velocity.\\ For every photon \psp\til solves the Kepler equation numerically and finds the corresponding eccentric anomaly $E$\cite{kramer04}.
If the pulsar companion has a strong gravitational field, the description in terms of Keplerian parameters
should be modified by including Post-Keplerian parameters. Since most of these parameters are very difficult
to measure and are known for only a few pulsars \cite{kramer04}, only the Keplerian model is currently  implemented in \psp.
For pulsars in binary orbits the arrival time corrections described in Eq.\ref{eq:bcorr} must
be modified to include additional terms:
\begin{equation}\label{eq:bcorr1}
t_{B} = t_{F} + \Delta t_{C} + \Delta t_{E} + \Delta t_{R} +
\Delta t_{S} + \Delta t_{RB} + \Delta t_{EB} + \Delta t_{SB}
\end{equation}
Since General Relativistic effects are not fully implemented in the current version of \psp\til
$\Delta t_{EB}$ (the binary \emph{Einstein delay}) and $\Delta_{SB}$ (the binary \emph{Shapiro delay}) are set to zero, and their magnitude is in the order of few $\mu$s \cite{kramer04}. The binary \emph{Roemer delay} $\Delta t_{RB}$ across the pulsar orbit
can be written in terms of the eccentric anomaly $E$ as:
\begin{equation}
\Delta_{RB}(t) = x(\cos E(t)-e)\sin \omega + x\sin E(t)\sqrt{1-e^{2}}\cos \omega
\end{equation}
where $x\equiv a_{p}\sin i$.
\section{Case studies of simulations}
In order to better understand the functionality of \psp\til and its use in testing the analysis tools as well as the LAT capabilities
for pulsar observations, some specific case studies are presented.
\subsection{PSR B1951+32}
The first example is a simulation of the light curve of PSR B1951+32, one of the faintest EGRET pulsars.
The simulation was performed by assuming one year of observation in scanning mode. To model the gamma-ray spectrum of PSR B1951+32 we used a power law with spectral index $\alpha$=1.74, a cutoff energy E$_{0}$=40 GeV and
a super-exponential cutoff index $b$=2 (spin parameters have been taken from EGRET observations).
The simulation included the diffuse gamma-ray background based on a model of cosmic rays and its interaction with the Interstellar Medium \cite{strong04} developed by the LAT collaboration.
\begin{figure}[ht]
\begin{center}
\includegraphics[width=0.7\columnwidth]{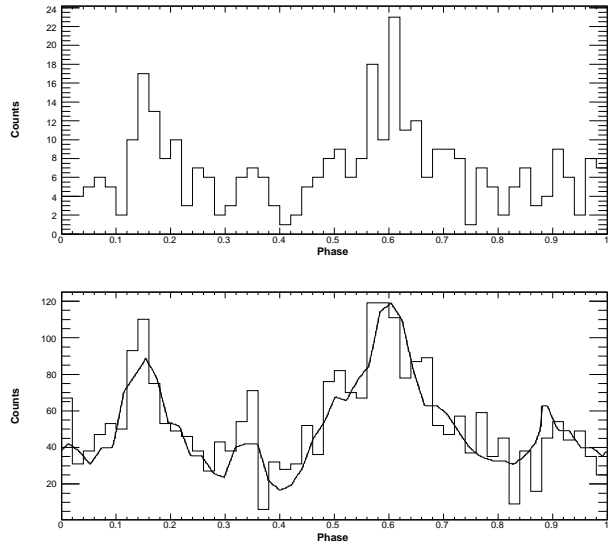}
\caption{Simulation of 1 year LAT observation of PSR B1951+32 in scanning mode compared with EGRET results. Top: 50 bin EGRET
light curve for photons around 1$^{\circ}$ from the pulsar position and energies above 100 MeV \cite{ramanamurthy95}. Bottom: 50 bin reconstructed light curve of LAT simulated data using the same region and energy range of EGRET data, compared with the simulated model (black line). Both light curves are background-subtracted.}\label{fig:fig4_1951}
\end{center}
\end{figure}
For comparison with the previous EGRET results, photons within a 1 degree circle around the source were selected. Simulated photons were then barycentered and phase-assigned using tools in the SAE. Following the same analysis steps performed on EGRET data, all the photons outside the phase intervals of the peaks
(0.12$<\phi<$0.22 and 0.48$<\phi<$0.74) were assigned to the background and subtracted.  The resulting
light curve is shown in Fig. \ref{fig:fig4_1951} together with the one from EGRET.\\
In this region EGRET collected 344 photons \cite{ramanamurthy95}. The statistics of the simulated LAT light curve are far superior to the EGRET observation, with the LAT light curve containing about 8 times as many gamma rays.
Since the LAT has a wide energy range, a greater number of photons is also expected because of the hard spectrum of this source. This example suggests the potential of the LAT to discover new gamma-ray pulsars fainter than PSR B1951+32.
\subsection{Vela pulsar}
We also present a case study that illustrates the capability of the LAT to measure pulsar spectral cutoffs in order to distinguish between Polar Cap and Outer Gap models, as outlined in Sec. \ref{sec:intro}. The Vela pulsar is the best candidate for spectral cutoff measurements, since it is the brightest source in the gamma-ray sky and its cutoff energy should be around a few GeV, well within the LAT energy range. For this reason we simulated 
1 year of LAT sky survey observations of the Vela pulsar using Polar Cap and Outer Gap models. The phase-averaged spectrum for Polar Cap is taken from the model of Daugherty and Harding 1996 \cite{daugherty96} while the Outer Gap from the model of Romani 1996 \cite{romani96}. The diffuse gamma-ray background is the same as was used for PSR B1951+32 simulations. The data were analyzed with the XSpec v12 package\footnote{$http://heasarc.gsfc.nasa.gov/docs/xanadu/xspec/$.} and a custom spectral model for fitting the super-exponential energy cutoff was used.
\begin{figure}[ht]
\begin{center}
\includegraphics[width=0.9\columnwidth]{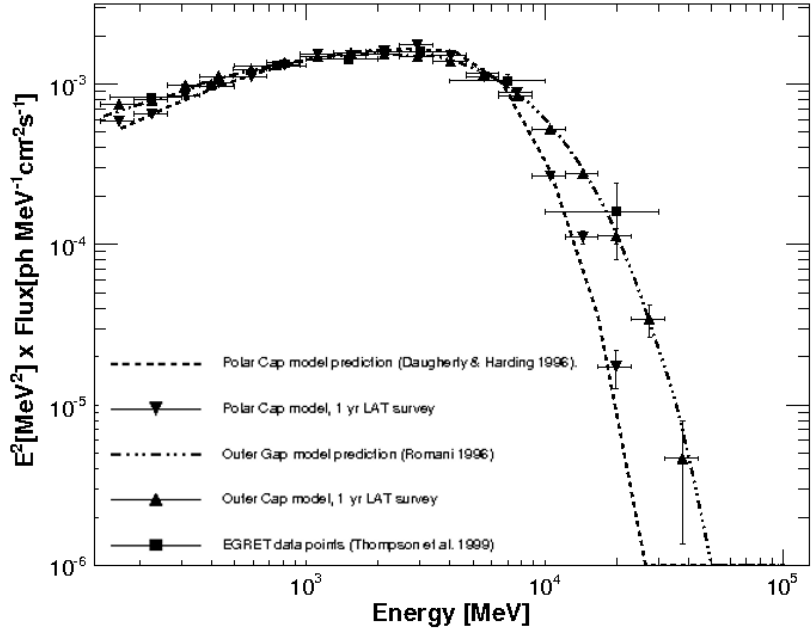}
\caption{Simulation of the Vela pulsar spectrum as observed in 1 year scanning mode. Polar Cap and Outer Gap spectrum are plotted together with the EGRET results to show the capability of the LAT to distinguish between them.}\label{fig:fig5_velasp}
\end{center}
\end{figure}
Fig. \ref{fig:fig5_velasp} shows the resulting spectral distributions for the LAT with the EGRET data superimposed. From this plot is appears that LAT is sensitive enough to constrain the spectral model for Vela.\\
It can be realistically estimated that the cutoff could be measured in a few months of observation in scanning mode, as it has been confirmed by Vela first observations by LAT\cite{abdo08c}. However, to have a clearer insight into the details of the emission mechanism, a longer exposure is needed. The most powerful analysis to study the emission mechanism is the phase-resolved spectral analysis, which should become feasible for the brightest pulsars within 1 year of data collection.\\
\subsection{Population Studies} 
\begin{figure}[ht] 
\begin{center} 
\includegraphics[width=0.95\columnwidth]{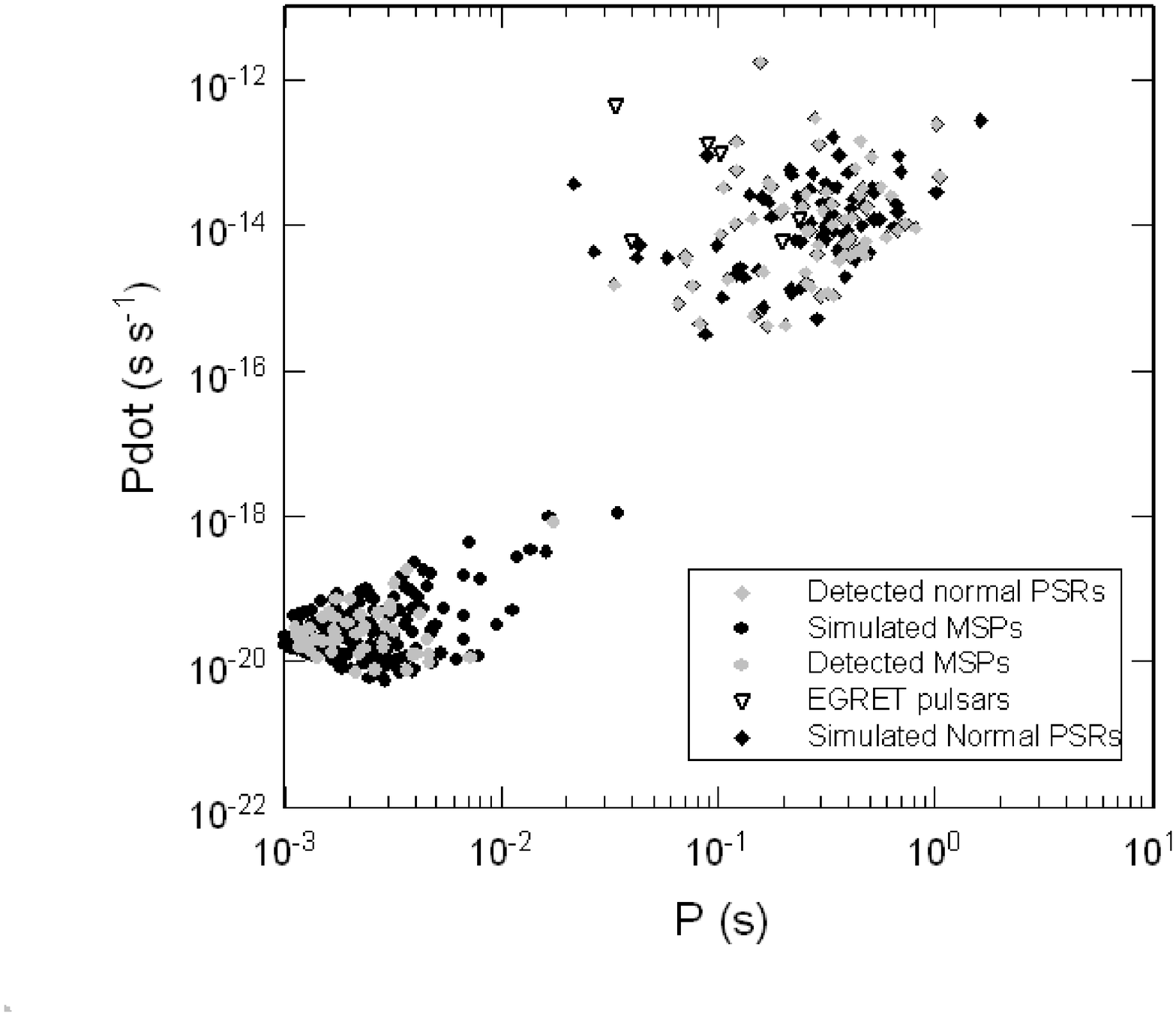} 
\caption{P-Pdot plot showing simulation of a population of 404 pulsars including different emission models, compared with the expected detections after 1 year (71 normal pulsars and 61 MSPs).}\label{fig:fig6_dc2} 
\end{center} 
\end{figure}

This example is based on the capability of simulating an entire pulsar population, with the goal of studying the LAT sensitivity \cite{razzano07} or understanding the LAT detection ratio between radio-loud and radio-quiet pulsars \cite{harding01}. 
We have produced large simulations based on a diversified and highly-detailed population built by using evolutionary synthesis codes. One example of such a population is shown in Fig. \ref{fig:fig6_dc2}, where a total of 404 pulsars down to a flux limit of 10$^{-9}$ph cm$^{-2}$s$^{-1}$ (E$>$100 MeV) was produced starting from the pulsar distribution provided by the population synthesis code described in \cite{gonthier02,harding05}. 
The population is composed of the 6 EGRET pulsars and 39 radio pulsars coincident with 3EG sources that were synthesized using the classical Polar Cap cascade scenario \cite{gonthier02}. 140 pulsars with Low Altitude Slot Gap emission \cite{muslimov03} were also simulated, by using a super exponential cutoff according to this model. 103 were found to be fainter in the radio and were classified as radio-quiet, while the others were classified as radio-loud. The model in Fig. \ref{fig:fig6_dc2} includes also 17 radio-loud and 212 radio-quiet millisecond pulsars (MSPs). According to the Polar Cap scenario, MSPs can produce gamma rays by Inverse-Compton induced cascades, even if they are below the Curvature Radiation pair death line \cite{harding05}.\\ 
In order to estimate the number of pulsars detected with the LAT, we run an automatic routine based on the \emph{Fermi} Science tools, obtaining a total of 71 normal pulsars and 61 MSPs  detected in 1 year. This result, that takes into account the detailed structure of the diffuse gamma-ray background, is compatible with the number of pulsars discovered discovered by \emph{Fermi} in its first months \cite{abdo09a}.\\ 
The algorithms in \psp~ have been optimized to keep simulation times short, and the simulation of large sample of pulsars have been used to estimate the computing time and precision.With a 2 GHz processor the simulation of 200 pulsars for an observation time of 1 month takes nearly two hours.
The precision that \psp~ can achieve has a direct impact on the CPU time. In order to have a good compromise between precision and CPU time, the configuration adopted for PulsarSpectrum has an accuracy of 50$\mu$s over the whole timing correction chain. This precision is considered satisfactory  for the purpose of this simulation but can be improved if necessary.
\subsection{Timing noise and blind search}
\begin{figure}[ht]
\begin{center}
\includegraphics[width=0.8\columnwidth]{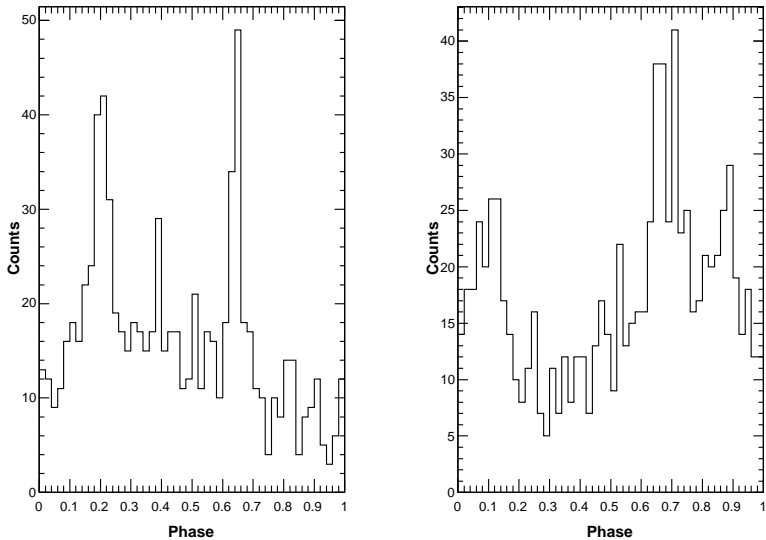}
\caption{Simulation of the effects of timing noise on a simulated pulsar over a month. Left: The pulsar is simulated without timing noise and phases are assigned using the exact values of $P_{0}$,$P_{1}$, and $P_{2}$. Right: The same pulsar simulated including timing noise and with phase assignment using only an approximation of $P_{0}$, and $P_{1}$ as in a typical blind search.}\label{fig:fig7_noisylc}
\end{center}
\end{figure}

\begin{figure}[ht]
\begin{center}
\includegraphics[width=1.0\columnwidth]{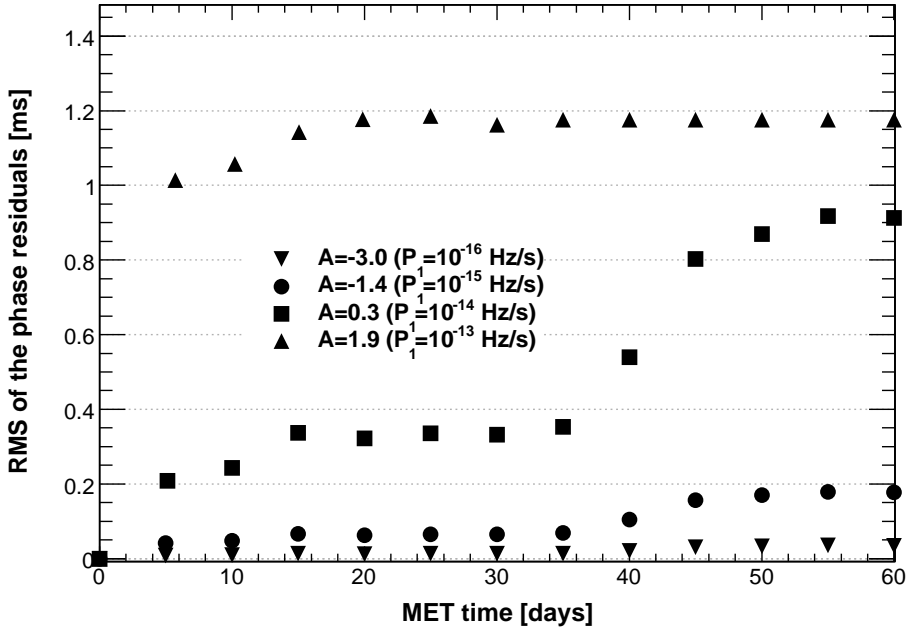}
\caption{Evolution of the RMS of the simulated timing noise with time for different values of P$_{1}$ corresponding to different values of activity $A$.}\label{fig:fig8_rmsvstime}
\end{center}
\end{figure}

The capability to simulate timing noise, which affects pulsar phase coherence over time (Sec. \ref{sec:tnoise}), allows a better understanding of the limits of the blind search for Radio Quiet pulsars. The first result of blind searches in the LAT data has been the discovery of the  CTA 1 supernova remnant\cite{abdo08a}.\\
The timing noise produces a distortion in the light curve, as is shown in Fig. \ref{fig:fig7_noisylc}, where an observation spanning 1 month is shown. With longer observation times the profile of the light curve continues to degrade until the pulsation is lost. 
The left panel shows a simulated pulsar located at CTA 1 with $P_{0}$=100 ms, $P_{1}$=10$^{-13}$ s s$^{-1}$, and $P_{2}$=9$\times10^{-24}$ Hz $^{-2}$ and no timing noise. The phases have been assigned using the exact values of $P_{0}$,$P_{1}$, and $P_{2}$ used for the simulation. 
The right panel shows a simulation of the same pulsar with timing noise, that has an activity parameter of $A\sim$1.89 (Eq. \ref{eq:tna_pdot}). The phase assignment has been carried out using approximate values of $P_{0}$, and $P_{1}$, as they would be obtained from a blind search. 
In Fig. \ref{fig:fig7_noisylc} a periodic structure is still visible but the significance is smaller, since the H-test statistics TS$_{H}$ change from TS$_{H}$=209 (left) to TS$_{H}$=125 (right).\\
A more detailed set of simulations has been produced to explore how the timing noise RMS evolves with time for different activity parameters derived using typical values of $P_{1}$ and keeping the same pulsar frequency of 10 Hz. We expect that higher activity parameters imply a more rapid increase of the timing noise RMS with time, as it is confirmed in Fig. \ref{fig:fig8_rmsvstime}.
An 0.1 ms or 1 ms RMS is compatible with typical RMS residuals provided from radio observations. It is clear from Fig. \ref{fig:fig8_rmsvstime} that a pulsar with a timing noise activity of 1.9 reaches an RMS value of 0.1 ms over few days while the same value is reached in about 2 months for smaller timing noise.
\section{Conclusions}
The first astronomical sources detected at gamma-ray energies were pulsars, but still many questions about their natures and emission mechanisms remain unanswered. $Fermi$ provides an enormous leap in capability for exploring pulsars and the LAT will be able to discover new gamma-ray pulsars, probing their nature with unprecedented detail.\\
Simulation is a powerful tool to study the response of the instrument and to check data analysis software. For this reason the LAT collaboration has produced a complete simulation package for the LAT detector, for different classes of sources and background, and a suite of software tools for analysis of gamma-ray data.\\
\psp, presented in this paper, is a new simulation package specific for gamma-ray pulsars. It has been demonstrated to be very flexible, accommodating several alternative models, and very accurate, taking into account the main timing effects on the arrival time of the photons. The motion of the spacecraft in the Solar System with relativistic corrections and timing noise due to unpredictable changes in pulsar period and phase have been included. It has been demonstrated that timing noise introduces a substantial limitation when searching for radio-quiet pulsars.
\psp\til has been, and will remain a very useful tool during the $Fermi$ mission as a way to compare real data with the predictions of different theoretical models.

\section*{Acknowledgments}
The $Fermi$ LAT Collaboration acknowledges generous ongoing support from a number of agencies and institutes that have supported both the development and the operation of the LAT as well as scientific data analysis.  These include the National Aeronautics and Space Administration and the Department of Energy in the United States, the Commissariat \`a l'Energie Atomique and the Centre National de la Recherche Scientifique / Institut National de Physique Nucl\'eaire et de Physique des Particules in France, the Agenzia Spaziale Italiana and the Istituto Nazionale di Fisica Nucleare in Italy, the Ministry of Education, Culture, Sports, Science and Technology (MEXT), High Energy Accelerator Research Organization (KEK) and Japan Aerospace Exploration Agency (JAXA) in Japan, and the K.~A.~Wallenberg Foundation, the Swedish Research Council and the Swedish National Space Board in Sweden.
Additional support for science analysis during the operations phase from the following agencies is also gratefully acknowledged: the Istituto Nazionale di Astrofisica in Italy and the K.~A.~Wallenberg Foundation in Sweden for providing a grant in support of a Royal Swedish Academy of Sciences Research fellowship for JC.

\end{document}